\input psfig.sty
\documentstyle{cupconf}

\ifoldfss
\else
  \ifnfssone
    \newmathalphabet{\mathit}
      \addtoversion{normal}{\mathit}{cmr}{m}{it}
      \addtoversion{bold}{\mathit}{cmr}{bx}{it}
    \newmathalphabet{\mathcal}
      \addtoversion{normal}{\mathcal}{cmsy}{m}{n}
    \else
    \ifnfsstwo
    \fi
  \fi
\fi

\def\solar{\ifmmode_{\mathord\odot}\else$_{\mathord\odot}$\fi~}
\def\hub{$H_0=100$ km s$^{-1}$ Mpc$^{-1}$, $q_0 = 0.5$}
\def\sprop{$S_{\nu} \propto \nu^{\alpha}$}
\def\gsim{\stackrel{>}{_\sim}}

\def\deg{\ifmmode $\setbox0=\hbox{$^{\circ}$}$^{\,\circ}
          \else    \setbox0=\hbox{$^{\circ}$}$^{\,\circ}$\fi\,}
\def\hexnumber#1{\ifcase#1 0\or1\or2\or3\or4\or5\or6\or7\or8\or9\or
 A\or B\or C\or D\or E\or F\fi }

\makeatletter
\ifx\CUP@mtlplain@loaded\undefined
\else
\fi
\makeatother
 \makeatletter
 \ifx\CUP@mtlplain@loaded\undefined
   \font\tenbmi=cmmib10 at 10pt
   \font\sevenbmi=cmmib10 at 7pt
   \font\fivebmi=cmmib10 at 5pt

   \newfam\bmifam
   \textfont\bmifam=\tenbmi
   \scriptfont\bmifam=\sevenbmi
   \scriptscriptfont\bmifam=\fivebmi
   
 \fi
 \makeatother
\ifnfsstwo

\fi
\ifnfssone

\fi
\ifoldfss

\fi

\mathchardef\varLambda="0103

\makeatletter
\ifx\CUP@mtlplain@loaded\undefined
\else
\fi
\makeatother
\makeatletter
\ifx\CUP@mtlplain@loaded\undefined
  \font\tenbms=cmbsy10
  \font\sevenbms=cmbsy10 at 7pt
  \font\fivebms=cmbsy10 at 5pt
  \newfam\bmsfam
  \textfont\bmsfam=\tenbms
  \scriptfont\bmsfam=\sevenbms
  \scriptscriptfont\bmsfam=\fivebms

  \edef\bsy@{\hexnumber\bmsfam}
  \mathchardef\bnabla="0\bsy@72
\fi
\makeatother
\title[Multi-Frequency Monitoring of 3C\,273]{  
From Centimeter to Millimeter  Wavelengths: A High Angular Resolution Study of 3C\,273
}

\author[T.P.Krichbaum et al.]{%
T\ls H\ls O\ls M\ls A\ls S\ns  P.\ns K\ls R\ls I\ls C\ls H\ls B\ls A\ls U\ls M\ls , A.\ns  W\ls I\ls T\ls Z\ls E\ls L\ls , a\ls n\ls d\ns J.\ns A.\ns  Z\ls E\ls N\ls S\ls U\ls S}

\affiliation{
Max-Planck-Institut f\"ur Radioastronomie, Auf dem H\"ugel 69, 53121 Bonn, Germany
} 

\begin{document}
\ifnfssone
\else
  \ifnfsstwo
  \else
    \ifoldfss
      \let\mathcal\cal
      \let\mathrm\rm
      \let\mathsf\sf
    \fi
  \fi
\fi

\maketitle

\begin{abstract}
We monitored 3C\,273 with VLBI at 15--86\,GHz since 1990. We discuss component trajectories,
opacity effects, a rotating jet, and outburst-ejection relations from Gamma-ray to radio bands.
\end{abstract}

\firstsection % if your document starts with a section,
\section{Introduction}
In radio-interferometry, the angular resolution can be improved either by increasing
the antenna separation or by observing at shorter wavelengths.
Antenna separations beyond the Earth's diameter lead to VLBI with
satellites, which presently is possible at cm-wavelengths (VSOP).
Complementary to this, ground-based VLBI observations at millimeter wavelengths (mm-VLBI)
provide even higher angular resolution (tens of micro-arcseconds), 
and facilitate imaging of compact structures, which are self-absorbed (opaque) and therefore 
not directly observable at the longer cm-wavelengths.

For the quasar 3C\,273 ($z=0.158$), the mm-VLBI observations 
provide images with an angular resolution of up to
$50$\,$\mu$as (1\,$\mu$as $= 10^{-6}$\,arcsec) at 86\,GHz. This corresponds to a spatial scale
of $\sim 1000$ Schwarzschild radii for a $10^9$\,M\solar black hole.
The small observing beam also allows to accurately determine positions of jet components
and facilitates to trace the bent jet
closer to its origin than before. This facilitates detailed studies of the jet structure
and kinematics near the nucleus, in particular with regard to the broad-band (radio to Gamma-ray)
flux density variability,
the injection of material (plasma) at the jet-base, and the birth of new `VLBI components'.

Here we summarize some new results from a multi epoch (1990 -- 1997) study at high observing frequencies
(15, 22, 43 and 86\,GHz). More details are given in Krichbaum et al. 2000. 
\begin{figure}[p]
\vspace{1.0cm}
\begin{minipage}[t]{6.5cm}{
\psfig{figure=kri_fig1.epsi,width=5cm}
\caption{
3C\,273 observed at 22\,GHz (top), 43\,GHz (center),
and 86\,GHz (bottom) nearly at the same epochs of 1995.15 (22 and 43\,GHz) and
1995.18 (86\,GHz). Data at 22 \& 43\,GHz are from A. Marscher (priv. com.). Contour levels are
-0.5, 0.5, 1, 2, 5, 10, 15, 30, 50, 70, and
90\,\% of the peak flux density of 3.0 (top), 5.4 (center), and 4.7\,Jy/beam (bottom).
For the 22\,GHz map, the 0.5\,\% contour is omitted. All maps are restored
with a beam of $0.4 ~{\rm x}~ 0.15$\,mas in size,
oriented at $\rm{pa}=0 \deg$. The maps are arbitrarily centered on the
eastern component (the core), the dashed lines guide the eye and help to identify
corresponding jet components in the three maps. 
}
}
\end{minipage}
~~
\begin{minipage}[t]{6.5cm}{
\vspace{-14cm}
\psfig{figure=kri_fig2.epsi,width=6cm,angle=-90}
\caption{Spectral index variations along the jet. For 1997 (circles, solid lines),
the spectral index gradient is calculated directly from the intensity profiles of the
maps at 15 and 86\,GHz. For 1995 (squares, dashed line), the spectral indices were
derived from Gaussian component model fits at 22 and 86\,GHz. We note that 
during the time interval 1995 -- 1997, different jet components occupied this jet region.
The spectral profile along the jet, however, did not change significantly.
}
\vspace{1cm}
\psfig{figure=kri_fig3.epsi,width=6cm,angle=-90}
\caption{The mean ridge line of the jet of 3C\,273 at 15\,GHz. Symbols denote
for observations at different epochs: 1995.54 (circles), 1996.94 (squares), 1997.04 (diamonds),
and 1997.19 (triangles). The offset is calculated relative to a straight
line, which defines the mean jet axis for $r < 20$\,mas. The straight line
is oriented along $\rm{pa}=240\deg$. We note the motion of the ridge line to the
right (south-west). This corresponds to an apparent pattern velocity of $\beta_{app} \simeq 4.2$.
}
}
\end{minipage}
\end{figure}
\section{Results}
At sub-milliarcsecond resolution, 3C\,273 shows a one sided core-jet structure of several
milliarcseconds length. The jet breaks up into multiple VLBI components, which -- when
represented by Gaussian components -- seem to separate at apparent superluminal
speeds from the stationary assumed VLBI core. The cross-identification of the model-fit components,
seen at different times and epochs, is facilitated by small ($< 0.2$\,mas), and to first
order negligible, opacity shifts of the component positions relative to the VLBI-core. 
Quasi-simultaneous data sets (cf. Figure 1) demonstrate convincingly the reliability of the component 
identification, which results in a kinematic scenario, in which all detected jet components
(C6 -- C18) move steadily (without `jumps' in position) away from the core (Figure 4). For the components
with enough data points at small ($<2$\,mas) and large ($> 2$\,mas) core separations, quadratic fits
to the radial motion r(t) (but also for x(t) --right ascension, and y(t) --declination) represent
the observations much better than linear fits.
Thus the components seem to accelerate as they move out. The velocities range typically from
$\beta_{app} =4 -8$ (for \hub).

From quasi-simultaneously obtained maps in 1995 (22/86\,GHz) and 1997 (15/86\,GHz),
we derived the spectral index gradient along the jet (see Figure 2). The spectrum oscillates
between $-1.0 \leq \alpha \leq +0.5$ (\sprop).  Most noteworthy, the spectral gradients did not change
significantly over the 2 year time period, although different jet components occupied 
this jet region. Thus, the geometrical (eg. relativistic aberration) and/or the physical environment
(pressure, density, B-field) in the jet must determine the observed properties of the VLBI components.
Hence, the latter do not form `physical entities', but seem to react to the physical 
conditions of the jet fluid. 

Further evidence supporting the impression that fluid dynamics determines the jet comes from a study
of the mean jet axis and the transverse width of the jet. Both oscillate
quasi-sinusoidally on mas-scales. At 15\,GHz, the variation of the ridge-line with time 
could be determined accurately from 4 high dynamic range maps, obtained during 1995 -- 1997 with the VLBA
(see Figure 3). The maxima and minima of the ridge-line are systematically displaced. This `longitudinal'
shift suggests motion with a pattern speed of $\beta_{app} =4.2$. The sinusoidal curvature of
the jet axis, however, is more indicative for jet rotation rather than for lateral displacements. Helical
Kelvin-Helmholtz instabilities propagating in the jet sheath could mimic such rotation, which, when
seen in projection, would explain this observation. 
\begin{figure}[t]
\vspace{0.5cm}
%\centerline{\psfig{figure=idall12.xvgr.epsi,width=10cm,angle=-90}}
\centerline{\psfig{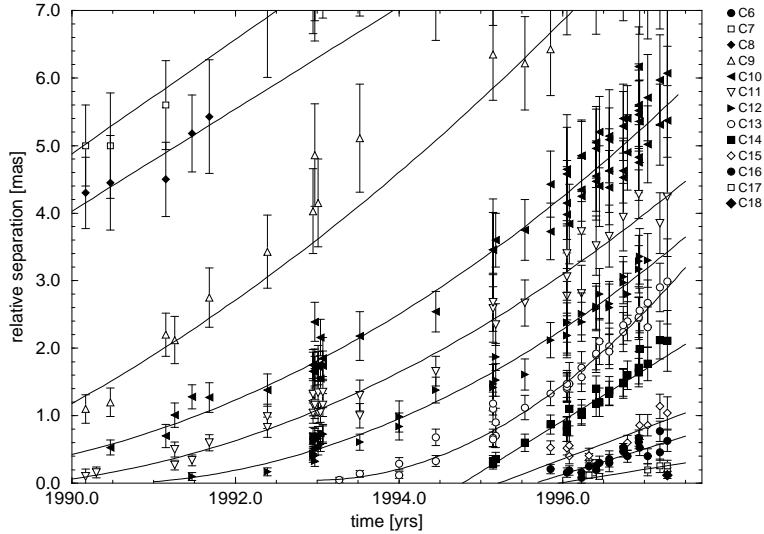}}
\caption{Relative core separations r(t) for the components C6 -- C18.
The legend on the right identifies symbols with VLBI components. The lines
are least square fits to the data.
}
\end{figure}

In Figure 5 we regard the x--y paths for the individual jet components. The lines represent quadratic
fits to the measured component positions. The figure on top shows all data, below only the inner
2\,mas region is shown. On the larger scale ($r < 20$\,mas), the components move along slightly 
curved trajectories. The coexistence of concave and convex shapes suggests again helical motion. 
The paths of C10 and C14 seem to define the northern and southern boundary of a conically expanding jet.  
In the central 2\,mas region (Figure 5, bottom), the trajectories are systematically displaced: 
between 1988 and 1995 the paths continuously rotate south (C11 -- C14), after 1995 they rotate 
back to the north (C15 -- C17). This rotation of the inner jet is displayed in a different manner
also in Figure 7 (bottom, left panel). Here we plot the position angle of the inner jet (derived
from a linear fit to the jet axis for $r < 0.5$\,mas) versus the back-extrapolated ejection
time $t_0$ (from Figure 4) of the VLBI components. This `ejection angle' varies 
periodically (period $T \simeq 15 -16$\,yr), with an amplitude of $\sim 30-40$\deg.
In the panel above, we plot the ejection velocity versus $t_0$. While the variation of the 
position angle basically agrees with the prediction from the precessing beam model 
proposed by Abraham \& Romero (1999) (dashed line in Figure 7),
the variation of the apparent speed is not well fitted. It looks, as if the apparent velocity
varies 2 times faster than the ejection angle. We also note that over the 37 years of observing time,
the apparent speed seems to decrease with a rate of $d\beta_{app}/dt= -(0.16 \pm 0.04)$\,yr$^{-1}$.
We tentatively suggest that the true precessing period therefore is $\gsim 150$\,yr, and
that the $\sim 8$ and $\sim 16$\,yr periods are superimposed on this. At present it is unclear,
whether the faster time scales result from some sort of nutation, or reflect time scales 
typical for helical KH-instabilities. The fact that the variation of the ejection velocity does 
not correlate with the flux density variability at mm and cm-wavelengths, but does correlate with 
the optical flux (Figure 7, right), seems to favor a jet intrinsic interpretation, rather 
than a geometrical origin. 

\hspace*{0.3cm} Although opacity shifts in the jet are small, they appear very clearly in the innermost part
of the jet. In Figure 6, we plot the orientation of the inner jet 
versus core separation. Near the core, at $r < 2-3$\,mas, the bending of the jet depends
on frequency, with a position angle offset of up to $\simeq 15$\deg between 15 and 86\,GHz. 
This suggests that near the core the curvature is 
frequency dependent and increases towards the core. At larger separations ($r > 3$\,mas), the 
offsets disappear. The most plausible explanation for this rests upon
opacity effects in a bend jet flow. With increasing frequency, the observer's line 
of sight penetrates deeper into the jet sheath. This leads to the apparent stratification and 
raises the question, if the jet speed also changes with frequency (velocity stratification).
The available data dot not yet give us a definitive answer on this.

For many AGN, a correlation between flux density variability and ejection of VLBI components 
is suggested. The large number of identified jet components and the long period of monitoring 
allows to investigate such outburst-ejection relations in 3C\,273 more systematically, than in 
other blazars. The each other complementing combination of high angular resolution from  mm-VLBI, and 
high sensitivity from cm-VLBI, allowed to identify 13 jet components (C6 -- C18)
and trace their motion back to their ejection from the VLBI core. The typical error in the determination
of the ejection times $t_0$ for each VLBI component ranges between 0.2 -- 0.5\,yr. In Figure 8 (left), 
we plot $t_0$ together with the light-curves at 22, 86, and 230\,GHz. We also add the Gamma-ray
detections of 3C\,273 from EGRET. T\"urler et al. 1999 have decomposed the multifrequency (mm- to cm-band)
light-curves into a sequence of individual flares,
and determined the time of onset for each sub-mm/mm-flare ($t_0^{\rm mm}$). Instead of looking at the
individual light-curves, we plot in Figure 8 (right) these onsets (open squares) together with the
VLBI ejection time ($t_0$) and the Gamma-ray fluxes. For each beginning of a mm-flare, we find that also
a new VLBI component was detected. Although the time sampling of the Gamma-ray data is quite coarse,
a relation between component ejection and high Gamma-ray flux appears very likely (note that detection 
at  Gamma-rays already means higher than usual $\gamma-$brightness). From a more detailed analysis
(Krichbaum et al. 2000) we obtain for the time lag between component ejection and onset of
a mm-flare: $t_0 - t_0^{\rm mm} = 0.1 \pm 0.2$\,yr. If we assume that the observed peaks in the Gamma-ray
light-curve (Figure 8) are located near the times $t_0^\gamma$ of flux density maxima, we 
obtain  $t_0^\gamma -  t_0^{\rm mm} = 0.3 \pm 0.3$\,yr. Although the Gamma-ray variability
may be faster, this result is fully consistent with
with the observations of Valtaoja \& Ter\"asranta (1996), who find (from a statistical analysis of
AGN) enhanced Gamma-ray fluxes mainly during the rising phase of millimeter flares.
We therefore suggest the following tentative sequence of events:
$t_0^{\rm mm} \leq  t_0 \leq t_0^{\gamma}$ -- the onset of a millimeter
flare is followed by the ejection of a new VLBI component and,
either simultaneously or slightly time-delayed, an increase of the Gamma-ray flux.
If we look only at the VLBI components, which were ejected close to the main maxima of the 
Gamma-ray light-curve in Figure 8, we obtain time lags of $t_0^{\gamma} - t_0$
of $\leq 0.5$\,yr for C12, $\leq 0.9$\,yr for C13, $\leq 0.2$\,yr for
C16 and $\leq 0.1$\,yr for C18. In all cases the Gamma-rays seem to
peak a little later than the time of component ejection. With $\beta_{app} \simeq 4$
near the core, the Gamma-rays would then escape at a radius $r_\gamma \leq 0.1$\,mas.
This corresponds to $r_\gamma \leq 2000$ Schwarzschild radii (for a $10^9$ M\solar black hole) 
or $\leq 6 \cdot 10^{17}$\,cm, consistent with theoretical expectations, in which
Gamma-rays escape the horizon of photon-photon pair production at separations  
of a few hundred to a few thousand Schwarzschild radii.\\
Obviously, more densely sampled VLBI- and Gamma-ray data would be needed to
decide, if the Gamma-rays originate from the Synchrotron-Self Compton process in the jet, or if the
seed photons for the Compton collision come from an external radiation field (eg. the BLR).

\begin{figure}
\begin{minipage}[t]{6.5cm}{
\psfig{figure=kri_fig5a.epsi,width=6cm,angle=-90}
\psfig{figure=kri_fig5b.epsi,width=6cm,angle=-90}
\caption{Projection of the component paths on the sky plane. Positions are
relative to the VLBI core, which is located at (x,y)=(0,0). The x-axis gives the
relative offset from the core in Right Ascension, the y-axis in Declination. Top: Trajectories
for components C6 -- C17. Bottom: Trajectories only for the central
2 mas region with components C9 - C17.} 
}
\end{minipage}
~~
\begin{minipage}[t]{6.5cm}{
\psfig{figure=kri_fig6.epsi,width=6cm,angle=-90}
\caption{The position angles of all VLBI components
plotted versus core separation. To reduce the scatter, a running mean
has been applied. Symbols denote different frequencies: 15 GHz filled
squares, 22 GHz open diamonds, 43 GHz filled circles, 86 GHz open triangles. At $r < 1$
mas, the position angles differ systematically with increasing frequency. Near the core, the 
position angle differences reaches $15\deg$.}
}
\end{minipage}
\end{figure}

\begin{figure}[t]
\vspace{1cm}
\begin{minipage}[t]{6.5cm}{
\psfig{figure=kri_fig7a.epsi,width=6cm,angle=-90}
}
\end{minipage}
~~
\begin{minipage}[t]{7.5cm}{
\psfig{figure=kri_fig7b.epsi,width=6.85cm,angle=-90}
}
\end{minipage}
\caption{Left: Apparent velocity $\beta_{app}$ (top)
and position angle (bottom) of the jet components near the core
plotted versus time of ejection $t_0$. On top, open circles show
the velocity after removal of an
overall slope of $d\beta_{app}/dt= -(0.16 \pm 0.04)$\,yr$^{-1}$.
Superimposed in both figures is the precessing beam model of
Abraham \& Romero, 1999 (long dashed line).
Right: $\beta_{app}$ (filled diamonds) plotted versus $t_0$.
Open diamonds show $\beta_{app}$ after removal of the
overall slope. Superimposed to the velocity is the optical V-Band light-curve
(from T\"urler et al. 1999). The optical flux density is in arbitrary units.
Note that due to the limited sampling prior to $\sim 1985$, ejection times and 
velocities are more accurately determined after this date.
}
%\end{figure}
%
%\begin{figure}[b]
\vspace{2cm}
\begin{minipage}[t]{6.5cm}{
\psfig{figure=kri_fig8a.epsi,width=6.5cm,angle=-90} 
}
\end{minipage}
~~
\begin{minipage}[t]{6.5cm}{
\psfig{figure=kri_fig8b.epsi,width=6.68cm,angle=-90}
}
\end{minipage}
\caption{Left: Flux density variations at 230 GHz (filled diamonds), 86 GHz (open circles)
and 22 GHz (filled squares). Upward oriented triangles denote Gamma-ray fluxes from EGRET,
downward oriented triangles show upper limits to the Gamma-ray flux.
Flux densities are in Jansky, except for the Gamma-ray, which is in
arbitrary units. The extrapolated
ejection times of the VLBI components and their uncertainties
are indicated by filled circles with horizontal bars along the time axis.
Right: Broad band flux density activity and component ejection.
VLBI component ejection (open circles), Gamma-ray fluxes (triangles, downward oriented for
upper limits) and
onset times for the millimeter flares (open squares, from T\"urler et al. 1999). Labels
denote the VLBI component identification.
}
\end{figure}

\end{document}